\documentclass[11pt,twoside]{article}
\usepackage{asp2010}
\usepackage{natbib}

\resetcounters

\begin{document}

\title{Differences of the Solar Magnetic Activity Signature in Velocity and Intensity Helioseismic Observations}
\author{D. Salabert,$^{1,2}$ R.~A. Garc\'ia,$^1$ and A. Jim\'enez$^{3,4}$
\affil{$^1$Laboratoire AIM, CEA/DSM-CNRS, Universit\'e Paris 7 Diderot, IRFU/Sap, Centre de Saclay, F-91191 Gif-sur-Yvette, France\\
$^2$Laboratoire Lagrange, UMR7293, Universit\'e de Nice Sophia-Antipolis, CNRS, Observatoire de la C\^ote d'Azur, F-06304 Nice, France\\
$^3$Instituto de Astrof\'isica de Canarias, E-38200 La Laguna, Tenerife, Spain\\
$^4$Departamento de Astrof\'isica, Universidad de La Laguna, E-38206 La Laguna, Tenerife, Spain}}

\begin{abstract}
The high-quality, full-disk helioseismic observations continuously collected by the spectrophotometer GOLF and the three photometers VIRGO/SPMs onboard the SoHO spacecraft for 17 years now
(since April 11, 1996, apart from the SoHO ``vacations") are absolutely unique for the study of the interior of the Sun and its variability with magnetic activity. Here, we look at the differences
in the low-degree oscillation p-mode frequencies between radial velocity and intensity measurements taking into account all the known features of the {\it p}-mode profiles (e.g., the opposite peak
asymmetry), and of the power spectrum (e.g., the presence of the higher degrees $\ell=$ 4 and 5 in the signal). We show that the intensity frequencies are higher than the velocity frequencies during
the solar cycle with a clear temporal dependence. The response between the individual angular degrees is also different. Time delays are observed between the temporal variations in GOLF and VIRGO frequencies. 
Such analysis is important in order to put new constraints and to better understand the mechanisms
responsible for the temporal variations of the oscillation frequencies with the solar magnetic activity as well as their height dependences in the solar atmosphere. It is also important for the study of
the stellar magnetic activity using asteroseismic data.
\end{abstract}

\section{Observations}
We analyzed more than 6000 days of continuous observations collected by the space-based instruments Global Oscillations at Low Frequency 
\citep[GOLF;][]{DS_gabriel95} and Variability of Solar Irradiance and Gravity Oscillations \citep[VIRGO;][]{DS_froh95}
onboard the {\it Solar and Heliospheric Observatory} (SoHO) spacecraft. GOLF is a resonant scattering spectrophotometer measuring the Doppler 
wavelength shift - integrated over the solar surface - in the D1 and D2
Fraunhofer sodium lines at 589.6 and 589.0 nm respectively. 
The GOLF velocity time series were obtained following \citet{DS_garcia05}.
VIRGO is composed of three sunphotometers (SPMs) at 402 nm (blue), 500 nm (green), and 862 nm (red). The analyzed period covers more than 16
years between 1996 and 2012. The overall duty cycles of the GOLF and VIRGO time series are 95.4\% and 95.2\% respectively. 

\section{Analysis}
The datasets were split into contiguous 365-day subseries, with a four-time overlap. The power spectrum of each subseries
was fitted by pair to extract the mode parameters \citep{DS_salabert07} using a standard likelihood maximization function (power spectrum with a $\chi^2$
with 2 degree of freedom (d.o.f.) statistics ). Each mode component was parameterized using an asymmetric lorentzian profile \citep{DS_nigam98}, the asymmetry being
assumed to be the same within a $\ell$ = 0 -- 2 and a $\ell$ =1 -- 3 pair of modes. The higher angular degrees $\ell$ = 4 and $\ell$ = 5 were included in the fitted model when visible 
(mainly between 2.5 and 3.3~mHz). Mean values of daily measurements of the 10.7-cm radio flux were
used as a proxy of the solar surface activity.

\section{Results}
\begin{figure}
\centering
\includegraphics[width=0.6\textwidth]{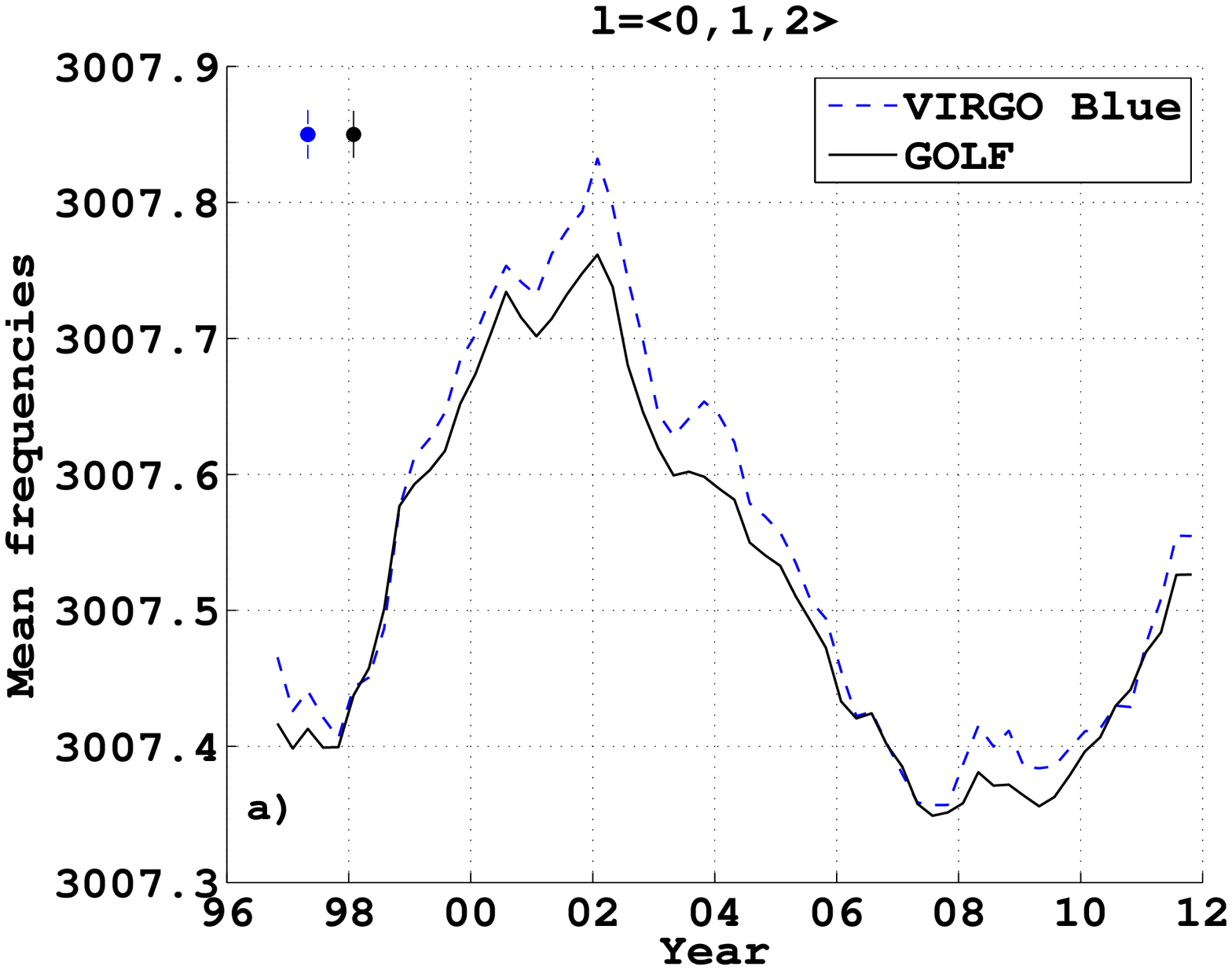}
\includegraphics[width=0.6\textwidth]{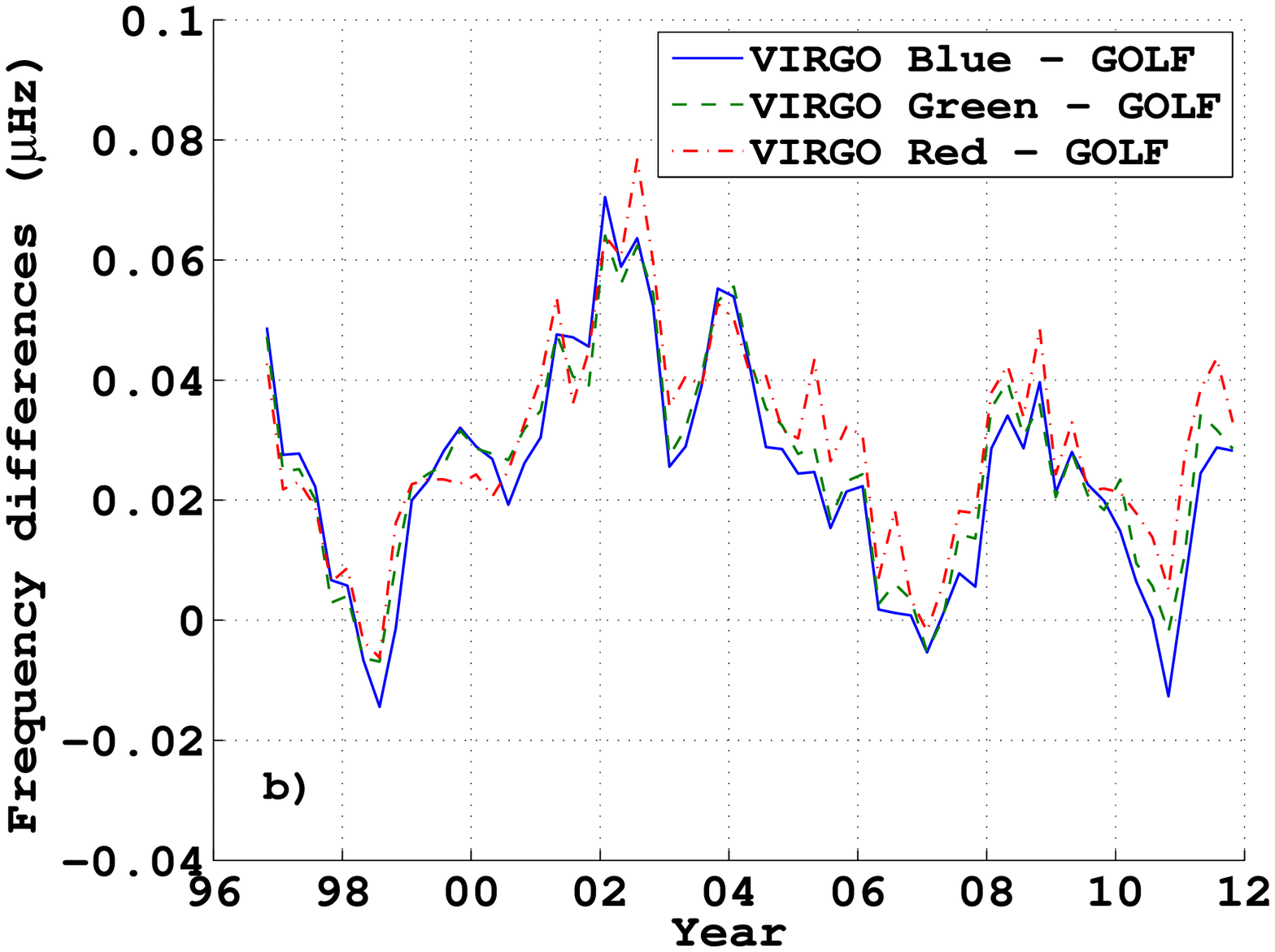}
\includegraphics[width=0.6\textwidth]{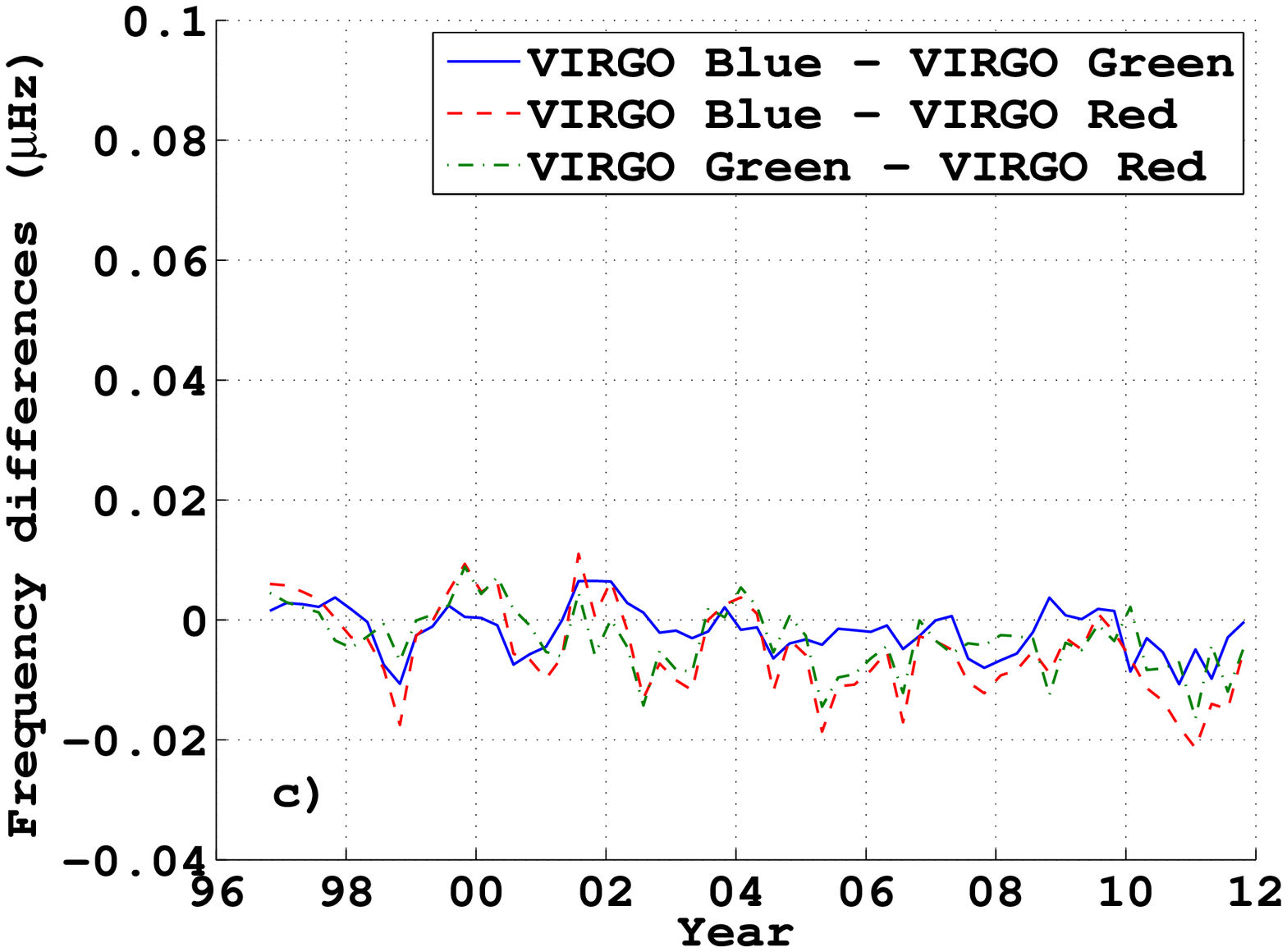}
\caption{(a) Temporal variations of the mean frequencies [2.5 -- 3.5mHz] of the $\ell$ = 0, 1, and 2 modes, $\langle \nu_{\ell=0,1,2}\rangle$, obtained with the velocity 
GOLF and intensity VIRGO/SPM blue channel measurements.
(b) Frequency differences between the VIRGO/SPMs and GOLF frequencies. (c) Same as (b) but between the VIRGO/SPMs.}
\label{fig:DS_fig1}
\end{figure}

The extracted mean frequencies between 2.5 and 3.5~mHz of the $\ell$ = 0, 1, and 2 modes, $\langle \nu_{\ell=0,1,2}\rangle$, are calculated on each 365-day segment 
for the different datasets.
The panel a) of Figure~\ref{fig:DS_fig1} shows $\langle \nu_{\ell=0,1,2}\rangle$ measured using the observations collected by GOLF and VIRGO (blue channel) as a 
function a time. The intensity frequencies are observed to be higher than the velocity frequencies, mainly during the solar maximum and minimum. Similar variations 
are observed between the three VIRGO/SPMs all over the solar cycle. Panels b) and c) show the frequency differences between the VIRGO/SPMs and GOLF, and between the 
three VIRGO/SPMs. We can see that intensity frequencies are systematically higher by about 30~nHz than the velocity frequencies over the solar cycle with a clear 
temporal dependence. This is in agreement with the ring-diagram analysis of the MDI velocity and continuum data performed by \citet{DS_tripathy09}, although they 
found larger systematic differences of the order of 5 -- 10~$\mu$Hz. 
This is in sharp contrast with the comparison of the {\it p}-mode frequencies measured from the three VIRGO/SPMs.

The cross correlations between the GOLF and the
VIRGO/SPMs frequencies, as well as the 10.7-cm radio flux, and
between the three VIRGO/SPMs are also computed. The obtained time delays are given in Table~\ref{table:DS_table1}. Negative time
delays of over 30 days are observed between frequencies from GOLF and VIRGO. Higher delays are also observed between GOLF and VIRGO when
compared to the 10.7-cm radio flux. Delays between the three VIRGO/SPMs are close to zero.

\begin{figure}
\centering
\includegraphics[width=0.6\textwidth]{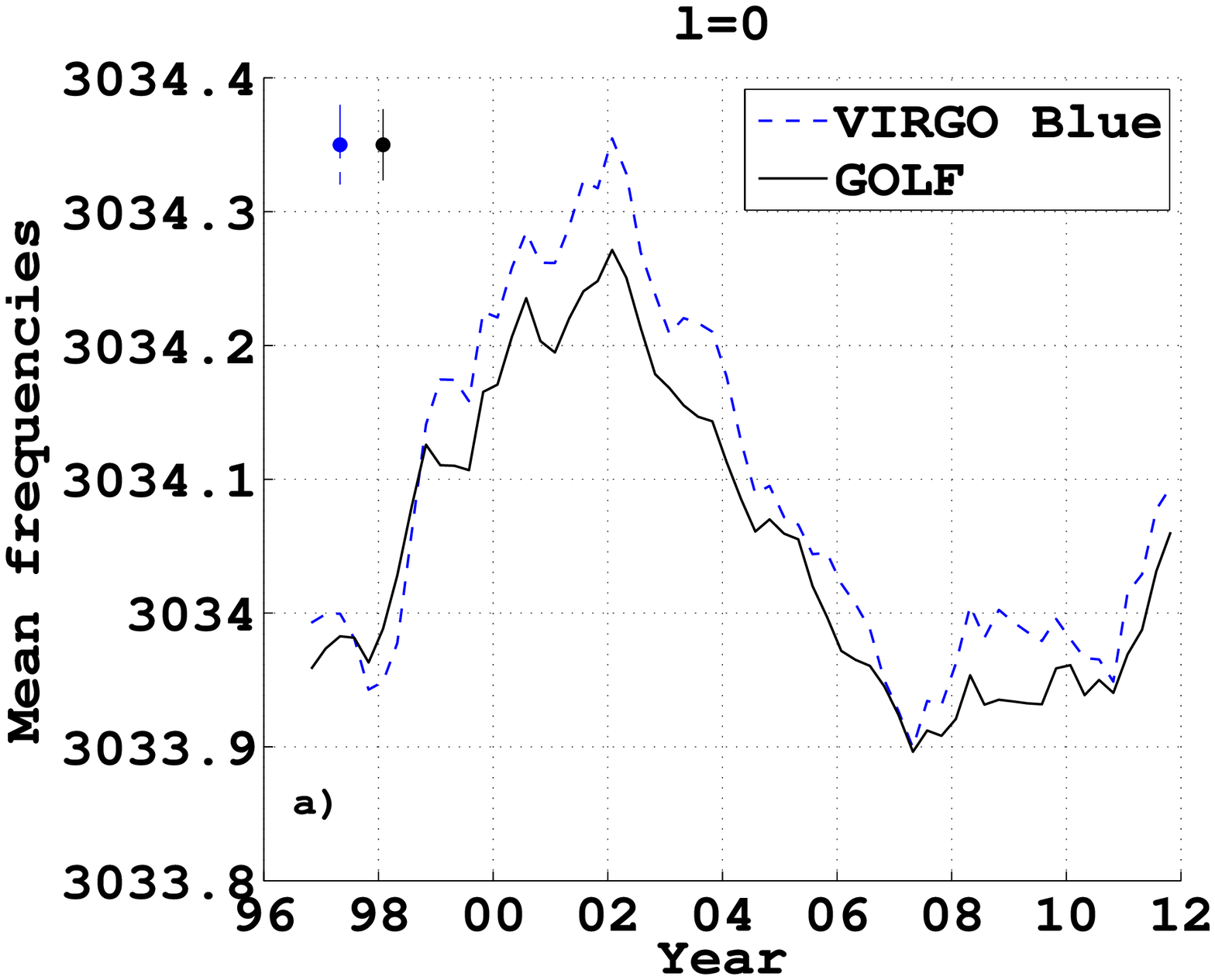}
\includegraphics[width=0.6\textwidth]{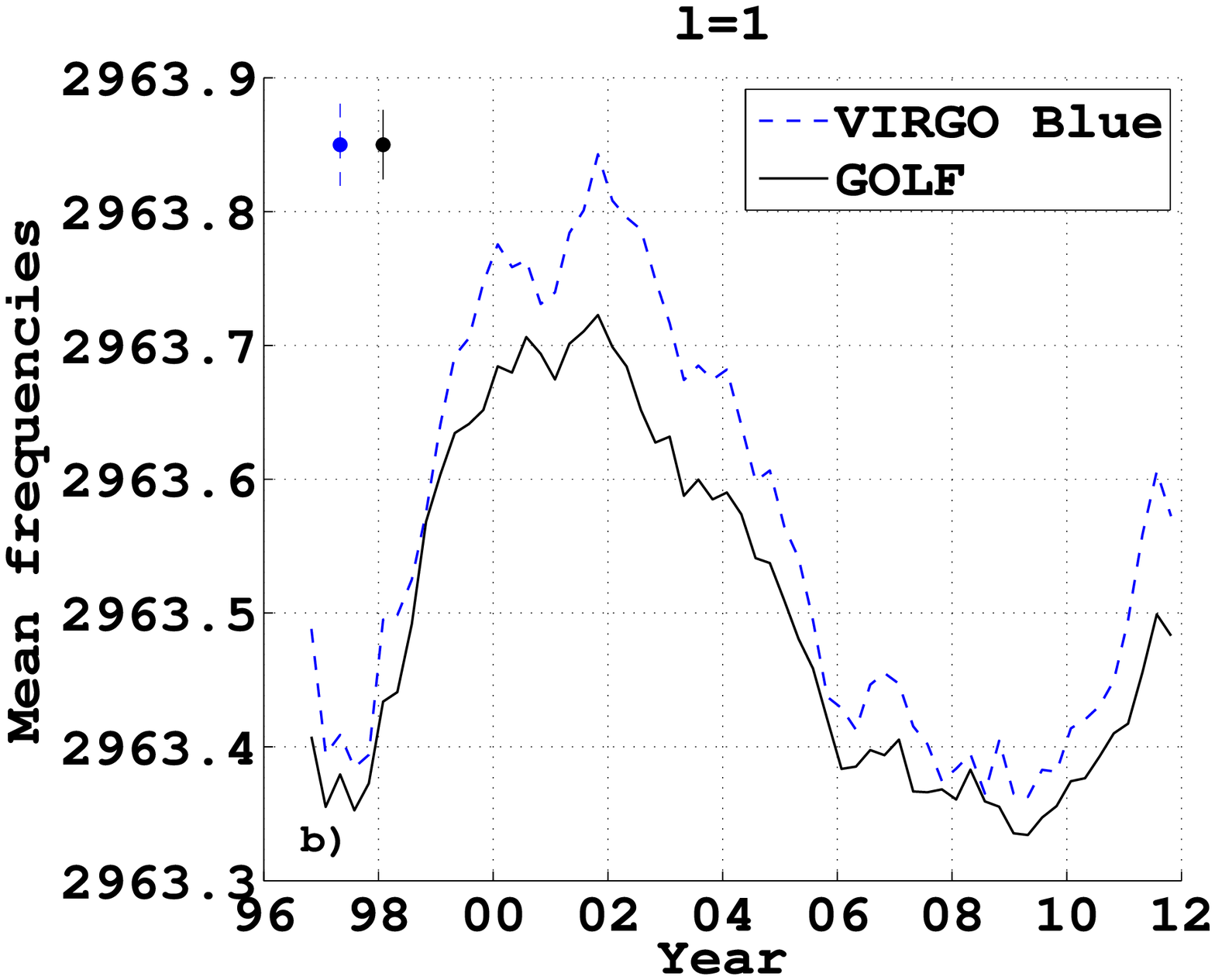}
\includegraphics[width=0.6\textwidth]{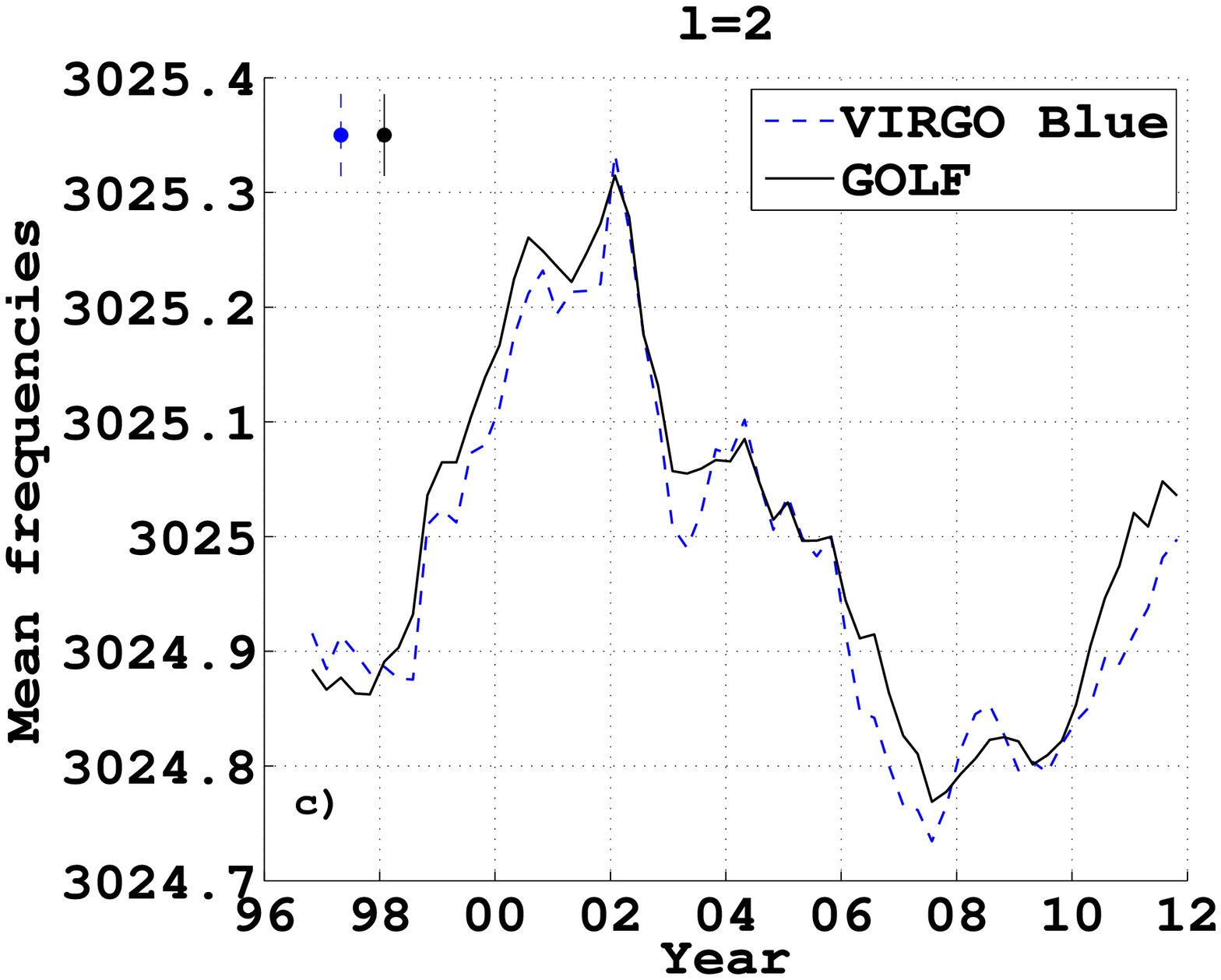}
\caption{Temporal variations of the mean frequencies [2.5 -- 3.5mHz] of the individual $\ell = 0$, 1, and 2 modes, 
$\langle \nu_{\ell=0}\rangle$,  $\langle \nu_{\ell=1}\rangle$, and $\langle \nu_{\ell=2}\rangle$,  observed with the velocity GOLF and intensity VIRGO/SPM blue channel measurements.}\label{fig:DS_fig2}
\end{figure}

Figure~\ref{fig:DS_fig2} shows the mean frequencies between 2.5 and 3.5~mHz, $\langle \nu_{\ell=0}\rangle$,  $\langle \nu_{\ell=1}\rangle$, and $\langle \nu_{\ell=2}\rangle$, 
for the angular degrees $\ell$ = 0, 1, and 2 individually. The main differences between intensity and velocity frequencies are at $\ell$ = 0 and $\ell$ = 1. Indeed, the intensity $\ell$ = 0
and 1 {\it p}-mode frequencies are observed to be higher than the velocity frequencies. The differences are larger during solar maximum, mainly in the case of the $\ell$ = 1 mode,
 where they can be over 100~nHz. On the other hand, the $\ell$ = 2 frequencies show almost no difference between intensity and velocity measurements.

\begin{table}[ht]
\caption{Time delays (in days) between velocity GOLF and intensity VIRGO/SPMs observations, with the 10.7-cm solar radio flux (F10.7), and between the three VIRGO channels.}
\label{table:DS_table1}
\smallskip
\begin{center}
{\small
\begin{tabular}{lc}
\tableline
\noalign{\smallskip}
Cross-correlation between: & Time delay (days) \\
\noalign{\smallskip}
\tableline
\noalign{\smallskip}
GOLF -- VIRGO Blue & $-38$  \\
GOLF -- VIRGO Green & $-33$ \\
GOLF -- VIRGO Red & $-37$ \\
\noalign{\smallskip}
\tableline
\noalign{\smallskip}
GOLF -- F10.7 & $~57$ \\
VIRGO Blue -- F10.7 & $~87$ \\
VIRGO Green -- F10.7 & $~87$ \\
VIRGO Red -- F10.7 & $~89$ \\
 \noalign{\smallskip}
\tableline
\noalign{\smallskip}
VIRGO Blue -- VIRGO Green & $-5$\\
VIRGO Blue -- VIRGO Red & $~~2$\\
VIRGO Red -- VIRGO Green & $~~3$\\
 \noalign{\smallskip}
\tableline
\end{tabular}
}
\end{center}
\end{table}

\section{Conclusions}
More than 6000 days of the radial velocity GOLF and photometric VIRGO/SPMs helioseismic observations are analyzed. 
Asymmetric lorentzian profiles are used to describe the solar acoustic oscillations in order to extract the {\it p}-mode 
parameters. Contiguous subseries of 365 days are used. The fitted velocity and intensity eigenfrequencies are compared. 
Systematic differences in the temporal variations of the p-mode frequencies up to 100~nHz are observed between the 
velocity GOLF and intensity VIRGO/SPMs  observations. These differences are more important for the $\ell$ = 0 and 1 modes 
and they appear to be larger during the solar maximum of the activity cycle. The $\ell$ = 2 frequencies show almost no difference. 
Preliminary study of the coherence functions between velocity and intensity data confirm these results; a detailed comparison 
is in progress. Also, cross-correlation analysis shows negative time lags between the GOLF and VIRGO frequencies of about 
30 days. Moreover, the temporal variations in GOLF and VIRGO frequencies  are observed to be {\it ahead} of 10.7-cm solar radio 
flux by about  60 and 90 days respectively. Time lags between the VIRGO channels are close to zero.

The VIRGO instrument observes the solar oscillations at the base of the photosphere, whereas GOLF observes a few hundred kilometers 
higher in the atmosphere \citep{DS_chano07}. The observed differences could arise from a height dependence of the {\it p}-mode frequency 
variations with solar activity. Moreover, are the differences observed at $\ell$ = 0  indirectly showing the Sun's radius 
varying? Also, these frequency differences might be larger in stars more active than the Sun. Nevertheless, there are few possible
 misleading origins for these differences between velocity and intensity observations. For instance, we cannot rule out probable 
interplays between the fitted peak asymmetry and the solar background and their variations with magnetic activity. Then, different 
realization noises  between intensity and velocity measurements might impact the fitted frequencies as well. Finally, the asymmetric 
lorentzian profile \citep{DS_nigam98} used to describe the p-mode oscillations might not be correct or one or more parameters might be 
missing in the description of the peak asymmetry.

\acknowledgements The GOLF and VIRGO instruments onboard SoHO are a
cooperative effort of many individuals, to whom we
are indebted. SoHO is a project of international
collaboration between ESA and NASA. The 10.7-cm radio flux data were
obtained from the National Geophysical Data Center.
This work has been partially supported by the
CNES/GOLF grant at the SAp CEA-Saclay.

\bibliography{salabert}

\end{document}